%% file: arttotal.tex
\documentstyle[12pt]{article}

\renewcommand{\O}{{\cal O}}
\renewcommand{\u}{u=\mbox{const.}}
\newcommand{\rr}{(\nabla r)^2}
\newcommand{\dr}{\triangle r\bigl|_{\mbox{\tiny AH}}}
\newcommand{\rH}{r_{\mbox{\tiny AH}}}
\newcommand{\vH}{v_{\mbox{\tiny AH}}}
\newcommand{\I}{{\cal I}^+}
\renewcommand{\|}{\biggl |_{{\cal I}^+}}
\begin{document}

\input arttxt.tex

\end{document}

%% file: arttxt.tex
\begin{center}
{\LARGE\bf
Backreaction of the Hawking radiation}
\end{center}
\begin{center}
{\bf G.A. Vilkovisky}
\end{center}
\begin{center}
Lebedev Physical Institute, and Lebedev
Research Center in Physics,\\ Leninsky Prospect 53, 119991 Moscow,
Russia.\\ E-mail: vilkov@sci.lebedev.ru \\
\end{center}
\vspace{3cm}
\begin{abstract}

Black holes create a vacuum matter charge to protect 
themselves from the quantum evaporation. A spherically
symmetric black hole having initially no matter charges
radiates away about 10\% of the initial mass and comes
to a state in which the vacuum-induced charge equals
the remaining mass.
\end{abstract}
\newpage

$$  $$

In refs. [1,2] a new approach is developed to the problem of
backreaction of the Hawking radiation. It brings to the solution.
The present paper is a report on the completion of this work.
For earlier studies and the background material see the book [3].

The Hawking radiation is a semiclassical effect, and so is its
backreaction [1]. If the collapsing matter has macroscopic
parameters, there is a region of the expectation-value spacetime
in which semiclassical theory is valid. This region is causally
complete [1] and covers the entire evolution of the black hole
from the macroscopic to the microscopic scale if the latter is
reached. The ultraviolet ignorance of semiclassical theory is
irrelevant to this region. In refs. [1,2] and the present paper,
the collapsing matter is assumed spherically symmetric, uncharged,
and having a compact spatial support. Then its only relevant
parameter is its mass $M$ which is also the ADM mass of the
expectation-value spacetime. The principal condition of validity
of the approach is $(\mu/M)\ll 1$ where $\mu$ is the Planckian
mass (1 in the absolute units). An observable that, in the units
of $M$, vanishes as $(\mu/M)\to 0$ is denoted as $\O$ and is a
microscopic quantity. An inequality of the form $X>|\O|$ assumes
any $\O$ and signifies that $X$ is a macroscopic quantity.

The key result [1] is that, in the semiclassical region of
spacetime, the expectation-value equations for the metric
close purely kinematically leaving the arbitrariness only in the
data functions. The data functions are two Bondi charges
appearing in the expansion of the metric at the future null
infinity $\I$:
\begin{equation}
\rr\|=1-\frac{2{\cal M}(u)}{r}+\frac{Q^2(u)}{r^2}+
O\left(\frac{1}{r^3}\right)\;.
\end{equation}
Here $r$ is the luminosity parameter of the radial light,
$u$ is the retarded time labelling the radial future light
cones, ${\cal M}(u)$ is the gravitational charge, and $Q(u)$
is a matter charge which can, for example, be the electric
charge. Any spherically symmetric metric is completely
specified by two local curvature invariants: $\rr$ and
$\mathop{\triangle}r$ where $\triangle$ is the D'Alembert
operator in the Lorentzian subspace. In the semiclassical
region, both invariants are expressed through the Bondi
charges.

On the other hand, the Bondi charges can be expressed through
the metric in the semiclassical region by calculating the
vacuum radiation against its background. As a result, the
Bondi charges get expressed through themselves, i.e., one
obtains closed equations for them and, thereby, for the
metric in the semiclassical region. The first stage of this
program: solving the kinematical equations for the metric in 
terms of the Bondi charges is accomplished in ref. [1].
The second stage: calculation of the vacuum radiation against
the thus obtained gravitational background is accomplished
in ref. [2]. The purpose of the present work is the solution
of the final equations.

Two normalizations of the retarded time $u$ figure in the
problem: $u^+$ and $u^-$. The $u^+$ is counted out by an observer
at infinity, and the $u^-$ is counted out by an early falling
observer [1]; $du^+/du^-$ is the red-shift factor. The $v$ below
is the advanced time labelling the radial past light cones and
counted out by an observer at infinity.

The self-consistent equations for the Bondi charges obtained in 
ref. [2] are of the form
\begin{eqnarray}
-\frac{d{\cal M}}{du^+}&\!=\!&\frac{\mu^2}{48\pi}\,\kappa^2
(1+\Gamma)\;,\\
\frac{dQ^2}{du^+}&\!=\!&\frac{\mu^2}{24\pi}\,\kappa\;,
\end{eqnarray}
\begin{equation}
\kappa=({\cal M}^2-Q^2)^{1/2}\left[{\cal M}+({\cal M}^2
-Q^2)^{1/2}\right]^{-2}
\end{equation}
where $\Gamma$ is expressed through ${\cal M}$ and $Q^2$ [2].
Here and below, the numerical factors are given for the vacuum
of spin-0 particles. Only the quantum s-mode contributes to
the flux of $Q^2$, and the fact that this flux is nonvanishing
is another key result [2]. The quantum modes with higher angular
momenta contribute only to the flux of ${\cal M}$ through the
term $\Gamma$ but $\Gamma$ is uniformly bounded and small:
$\Gamma\le(27/160)$. The spectral decomposition of the flux
(2) is Planckian with the time-dependent temperature [2]
\begin{equation}
kT=\mu^2\,\frac{\kappa(u)}{2\pi}\;.
\end{equation}

The data at $\I$, ${\cal M}$ and $Q^2$, are related to the data
at the apparent horizon (AH) as~[1]
\begin{equation}
\rH={\cal M}+({\cal M}^2-Q^2)^{1/2}\;,\qquad
\dr=2\kappa\;,
\end{equation}
and the equation of the apparent horizon in the coordinates
$(u,v)$, $v=\vH (u)$, is given by the law [1]
\begin{equation}
\frac{d\ln\beta}{du^+\hphantom{n}}=
\kappa\left(\frac{d\vH}{du^+}-1\right)\;,
\qquad \beta=-\frac{d\rH}{du^+}\;.
\end{equation}
The outgoing light rays $\u$ cross the AH twice, and there is
a point, $(u_0,v_0)$, at which the AH is tangent to an outgoing
light ray [1]. Equivalently, the AH has two branches with
the origin at $(u_0,v_0)$. The equations above pertain to the
second (later) branch, and their validity is limited to the range
\begin{equation}
u^+>u^+_0+O(M)
\end{equation}
in which a significant radiation occurs. In this range, the
red-shift factor is given by the expression [1]
\begin{equation}
\frac{du^+}{du^-}=\frac{1}{2\beta_0}\exp\biggl(
\int\limits^{u^+}_{u^+_0}\kappa\,du^+\biggr)\;,\qquad
\beta_0=\beta\Bigl|_{u=u_0}\;.
\end{equation}

Another and more important limitation on the validity of the
equations above stems from the fact that they were derived
under certain assumptions about the data functions [1].
These assumptions, deliberately valid at the beginning of
the radiation process, could cease being valid at some late
value of $u$, and it was envisaged that the solution for the
metric obtained in ref. [1] should then be cut off at this
value of $u$. For later $u$, it is no longer valid. Now, that
the data functions are obtained, it turns out that, of these
assumptions, the crucial one is
\begin{equation}
\kappa>|\O|\;.
\end{equation}

The equations above are easy to solve, and the result is this.
${\cal M}(u)$ decreases monotonically, and $Q^2(u)$ increases
monotonically from the instant $u_0$ at which their values (up
to $\O$) are
\begin{equation}
\makebox[14cm][l]{$\displaystyle u=u_0\,\colon
\qquad\qquad\qquad\qquad\quad
{\cal M}_0=M\;,\qquad Q_0{}^2=0$}
\end{equation}
to an instant $u_1$ at which ${\cal M}^2$ and $Q^2$ become equal:
\begin{equation}
\makebox[14cm][l]{$\displaystyle u=u_1\,\colon
\qquad\qquad\qquad\qquad\qquad\qquad
{\cal M}_1{}^2=Q_1{}^2\;.$}
\end{equation}
Approximately,
\begin{equation}
u^+_1-u^+_0=96\pi\,\frac{M^3}{\mu^2}\;,
\end{equation}
and
\begin{equation}
0{.}098<\frac{M-{\cal M}_1}{M}<0{.}112\;.
\end{equation}
Here the lower bound accounts for the contribution of the s-mode
alone. Thus only about 10\% of $M$ is radiated away by the instant
$u_1$. The temperature of radiation first grows but only up to
a maximum value which is slightly greater than $\mu^2/8\pi M$,
and next decreases down to zero. At $u=u_1$, the red shift
reaches its maximum:
\begin{equation}
\left.\frac{du^+}{du^-}\right|_1=\exp\Bigl(24\pi\,
\frac{{\cal M}_1{}^2}{\mu^2}\Bigr)\approx
\exp\Bigl(19{.}4\pi\,
\frac{M^2}{\mu^2}\Bigr)\;.
\end{equation}

Along the AH, $r$ decreases monotonically from the value 
$2{\cal M}_0$ to the value ${\cal M}_1$, and 
$(r\mathop{\triangle}r)$ decreases monotonically from 1 to 0.
Because of the limitations (8) and (10), eq. (7) can be used
only outside the $O(M)$ neighbourhoods of the end points
$u^+_0$ and $u^+_1$. There, the equation of the apparent
horizon is
\begin{equation}
{}\hphantom{\qquad\qquad\qquad\qquad{}}
\frac{d\vH}{du^+}=1+|\O|\;,\qquad
u^+_1-O(M)>u^+>u^+_0+O(M)\;.
\end{equation}
In the neighbourhoods of the end points, only unessential details
of the behaviour of the AH are unknown. At $u=u_0$, the AH is
null and tangent to the outgoing light ray $u=u_0$. At
$u=u_1$, the AH is null and tangent to the incoming light ray
$v=v_1$. Approximately,
\begin{equation}
v_1-v_0=96\pi\,\frac{M^3}{\mu^2}\;.
\end{equation}
The apparent horizon (the second branch) is shown in Fig. 1.
At $u=u_1$, the assumption (10) breaks down, and the solution
for later $u$ is presently unknown.

The Hawking process liberates more than half of the energy from
the black hole. Only a small part of it goes away in the form
of thermal radiation. If one forgets about the radiation and just
compares the initial state at $u=u_0$ and the final state at
$u=u_1$, then in the initial state all of the available energy
is in the black hole, and in the final state exactly one half.
Most of the liberated energy remains in the compact domain outside
the black hole in the form of the energy of a long-range field
whose source is the charge $Q$. The black hole manufactures
this charge from the vacuum to protect itself from the quantum
evaporation. It is surprising that the vacuum stress tensor
develops a macroscopic value and even more surprising that
this is the stress tensor of a long-range field. The most
intriguing question: if this field couples to any matter
remains without an answer. Only its energy-momentum tensor is 
available.

The state reached by the instant $u_1$ is not really final because
the black hole does not stabilize in this state. This is seen
from the fact that, at $u=u_1$, both $u^+$ and $du^+/du^-$
have finite values. Only in the region of weak field [1] does the
expectation-value metric appear as the Reissner-Nordstrom metric
with slowly varying parameters. In the neighbourhood and
interior of the AH, it is different [1]. The ${\cal M}(u)$
and $Q^2(u)$ come to the instant $u_1$ with vanishing derivatives
and can be continued as constants but this is not true of
$\rH$ and $\dr$. The $\rH$ continues decreasing while $\dr$
passes through zero and becomes negative. Therefore, the AH
continues through the point $(u_1,v_1)$ as shown in Fig. 1.
At $u=u_1$, the present solution ceases being valid {\it but
semiclassical theory does not}. It is a matter of generalizing
the solution, to learn what next.

It will be emphasized that the term "black hole" is used here
for the interior of the apparent horizon rather than of the
event horizon. No event horizon has thus far been found in
the solution. There is strictly speaking no black hole but
at each instant of evaporation there is an "instantaneous
black hole" [1]. At the beginning of the evaporation process,
this is the "classical black hole" that corresponds to the
sector $v_0<v<v_{\mbox{\scriptsize crit}}$ of the AH in Fig. 1.
The value $v_{\mbox{\scriptsize crit}}$ sets the limit to the
validity of the correspondence principle [1]. In ref. [1], it
has been expressed through the constant $\beta_0$ in eq. (9).
Now one is able to calculate it:
\begin{equation}
v_{\mbox{\scriptsize crit}}-v_0=4M\ln\frac{M^2}{\mu^2}+O(M)\;.
\end{equation}
Remarkably, one is able to calculate also the value of $u^+_0$:
\begin{equation}
u^+_0-u^+_{\mbox{\scriptsize early}}=
4M\ln\frac{M^2}{\mu^2}+O(M)\;,
\end{equation}
\begin{equation}
\left.\frac{du^+}{du^-}\right|_0=\mbox{const.}\,\frac{M^2}{\mu^2}\;.
\end{equation}
Here $u_{\mbox{\scriptsize early}}$ is any $u$ at which the red
shift is moderate:
\begin{equation}
\left.\frac{du^+}{du^-}\right|_{\mbox{\scriptsize early}}<
\frac{1}{|\O|}\;.
\end{equation}
Eq. (19) gives the time instant at which the "classical black hole"
forms, and eq. (18) gives its life time. In the classical geometry,
both are infinite. The expectation-value geometry contains
two characteristic time scales: the one in eq. (19) and the one
in eq. (13).

The present work was supported by the Italian Ministry for
Foreign Affairs via Centro Volta, and the Ministry of Education
of Japan via the Yukawa Institute for Theoretical Physics.
Special thanks to Luigi Cappiello and Roberto Pettorino, and
Masao Ninomiya and Mihoko Nojiri for their hospitality in Naples
and Kyoto respectively.
\vspace{15mm}
\hrule\smallskip
\begin{itemize}
\item[[1]] G.A. Vilkovisky, {\it Kinematics of Evaporating Black
Holes}, the accompanying paper.
\item[[2]] G.A. Vilkovisky, {\it Radiation Equations for Black
Holes}, the accompanying paper.
\item[[3]] V.P. Frolov and I.D. Novikov, {\it Black Hole Physics},
Kluwer, Dordrecht, 1998.
\end{itemize}

\newpage

\begin{center}
\section*{\bf Figure caption}
\end{center}

$$  $$

\begin{itemize}
\item[Fig.1.] The second branch of the apparent horizon in the
coordinates $u^+,v$.
\end{itemize}